\documentclass[aps,prl,reprint,superscriptaddress,letterpaper]{revtex4-1}
\usepackage[english]{babel}
\usepackage{ucs}
\usepackage[utf8x]{inputenc}
\usepackage{amsmath}
\usepackage{amsfonts}
\usepackage{amssymb}
\usepackage{graphicx}
\usepackage{bm}
\usepackage{bbm}
\usepackage{empheq}
\usepackage{hyperref}

\begin{document}

\title{Cavity-Mediated Entanglement Generation Via Landau-Zener Interferometry}
\author{C. M. Quintana}
\author{K. D. Petersson}
\author{L. W. McFaul}
\affiliation{Department of Physics, Princeton University, Princeton, New Jersey 08544, USA}
\author{S. J. Srinivasan}
\author{A. A. Houck}
\affiliation{Department of Electrical Engineering, Princeton University, Princeton, New Jersey 08544, USA}
\author{J. R. Petta}
\affiliation{Department of Physics, Princeton University, Princeton, New Jersey 08544, USA}
\affiliation{Princeton Institute for the Science and Technology of Materials (PRISM), Princeton University, Princeton,
New Jersey 08544, USA}

\date{\today}

\begin{abstract}
We demonstrate quantum control and entanglement generation using a Landau-Zener beam splitter formed by coupling two transmon qubits to a superconducting cavity. Single passage through the cavity-mediated qubit-qubit avoided crossing provides a direct test of the Landau-Zener transition formula. Consecutive sweeps result in Landau-Zener-St\"{u}ckelberg interference patterns, with a visibility that can be sensitively tuned by adjusting the level velocity through both the non-adiabatic and adiabatic regimes. Two-qubit state tomography indicates that a Bell state can be generated via a single passage, with a fidelity of 78\% limited by qubit relaxation.
\end{abstract}


\maketitle
Avoided crossings are common in the energy level spectra of many quantum systems. In general, an off-diagonal matrix element that couples two states with magnitude $\Delta$ leads to an avoided crossing of magnitude $2\Delta$ in the energy level spectrum \cite{Cohen-Tannoudji1977}. Adjusting an external parameter to sweep the system through such an avoided crossing leads to a non-adiabatic transition probability $P_\text{LZ}$, as first described by Landau, Zener, St\"{u}ckelberg, and Majorana \cite{Landau1932,Zener1932,Stuckelberg1932,Majorana1932}. Landau-Zener transition physics has been applied to atomic collisions, where the external parameter is interatomic distance, and to adiabatic rapid passage in nuclear magnetic resonance, where the parameter is a rapidly varying magnetic field \cite{Nakamura2002,Shevchenko2010}. Avoided crossings have recently been used as effective ``beam splitters" of quantum states in more controllable systems including single superconducting qubits \cite{Izmalkov2004,Oliver2005,Sillanpaa2006,Berns2008,Sun2010} and semiconductor double and triple quantum dots \cite{Petta2010,Gaudreau2011}.

Circuit quantum electrodynamics (cQED) uses a dispersive cavity-mediated interaction between two superconducting qubits placed in the same microwave resonator to generate an effective qubit-qubit coupling $g_{12}=\Delta/\hbar$ \cite{Blais2004,Wallraff2004,Majer2007,Simmonds2007}. The ``quantum bus" architecture has enabled the demonstration of two-qubit algorithms and three-qubit error correction \cite{DiCarlo2009,Reed2012}. In this Letter, we use Landau-Zener physics to control states in the single-excitation subspace of two transmon qubits coupled to a superconducting resonator. Single passage through the avoided crossing via fast magnetic flux sweeps provides a direct test of the asymptotic Landau-Zener formula. Double passage results in tunable Landau-Zener-St\"{u}ckelberg interference patterns, analogous to Mach-Zehnder interferometry. These experiments allow us to observe the full crossover from non-adiabatic to adiabatic transitions in a single controllable system. Finally, single passage through the avoided crossing is used to generate two-qubit entanglement with a relaxation-limited fidelity of 78\%, as verified by state tomography. Such a beam splitter of two-qubit states could potentially be of use in a quantum information processor in which tunable qubits utilize more than one bias point for computation.

\begin{figure}[t]
\begin{center}
		\includegraphics[width=\columnwidth]{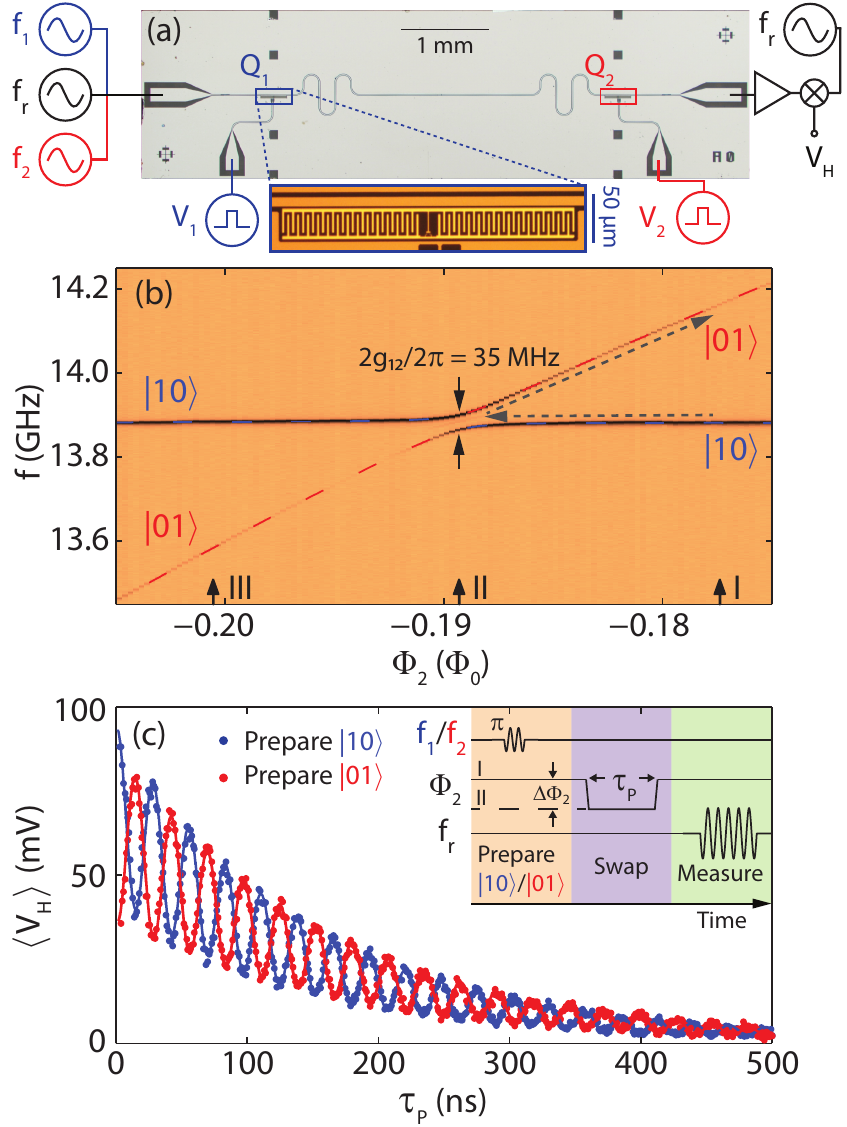}
\caption{\label{sc1} (color online) (a) Optical micrograph of the cQED device consisting of a $\lambda$/2 superconducting resonator coupled to two transmon qubits. Qubit $Q_1$ ($Q_2$) is controlled with flux bias voltage $V_1$ ($V_2$) and microwave tone $f_1$ ($f_2$). Qubit readout is performed by driving the resonator with a microwave tone $f_{\text{r}}$ and measuring the transmitted homodyne voltage $V_{\text{H}}$. (b) Low-power spectroscopy reveals a cavity-mediated avoided crossing between states $\left|10\right\rangle$ and $|01\rangle$. $\Phi_2$ denotes the applied magnetic flux through the split junction loop of $Q_2$, and $\Phi_0$ is the magnetic flux quantum. Qubit state readout is performed at point I. (c) A time-resolved cavity-mediated interaction is achieved by pulsing near the avoided crossing (point II) for a time $\tau_{\text{p}}$. The blue (red) points show $V_{\text{H}}$ as a function of $\tau_{\text{p}}$ when the system is initialized in $|10\rangle$ ($|01\rangle$) at point I. Solid lines are a guide to the eye.}
\end{center}	
\vspace{-0.5cm}
\end{figure}

Our cQED device, shown in Fig.\ \ref{sc1}(a), consists of two aluminum transmon qubits coupled to a niobium superconducting coplanar waveguide on a sapphire (Al$_2$O$_3$) substrate \cite{Koch2007,Dolan1977,Wallraff2004,Majer2007}. All measurements are performed in a cryogen-free dilution refrigerator with a base temperature of $\sim10$ mK. Each qubit is coupled to its own flux bias line, which we calibrate for both dc and ac crosstalk. Experimentally determined parameters include a fundamental cavity mode frequency $f_{\text{r}} = 8.795$ GHz, photon decay rate $\kappa/2\pi \approx 0.5$ MHz (for a cavity quality factor $Q \approx 17,000$), qubit-cavity couplings (between the first transmon transition and lowest cavity mode) $g_{1(2)}/2\pi = 183(185)$ MHz, transmon charging energies $E_{\text{C}1(2)}/h = 220(200)$ MHz, and transmon Josephson energies $E_{\text{J}1(2)}^\text{max}/h = 140(150)$ GHz.

The cavity-mediated qubit-qubit coupling is first probed through dispersive microwave spectroscopy, where transmission through the cavity is measured while a second microwave tone is swept across the transmon transition frequencies \cite{Wallraff2004,Schuster2005}. Transitions in the transmon qubits are detected as a dispersive shift in the cavity resonance, allowing qubit spectroscopy at frequencies much different from $f_{\text{r}}$. Figure \ref{sc1}(b) reveals a coupling strength of $2g_{12}/2\pi = 34.8$ MHz when the qubits are tuned into resonance at 13.9 GHz. A simple two-qubit gate is implemented by applying a $\pi$-pulse to one of the qubits at point I and then pulsing via the flux bias lines to the avoided crossing at point II for a variable amount of time $\tau_{\text{P}}$. The frequency of the resulting coherent oscillations, shown in Fig.\ \ref{sc1}(c), is consistent with the energy splitting obtained from spectroscopy. Qubit readout is performed using the Jaynes-Cummings nonlinearity, through which the strongly driven bare-cavity transmission is highly sensitive to the state of each qubit \cite{Reed2010}.

\begin{figure}[b]
\begin{center}
\includegraphics[width=\columnwidth]{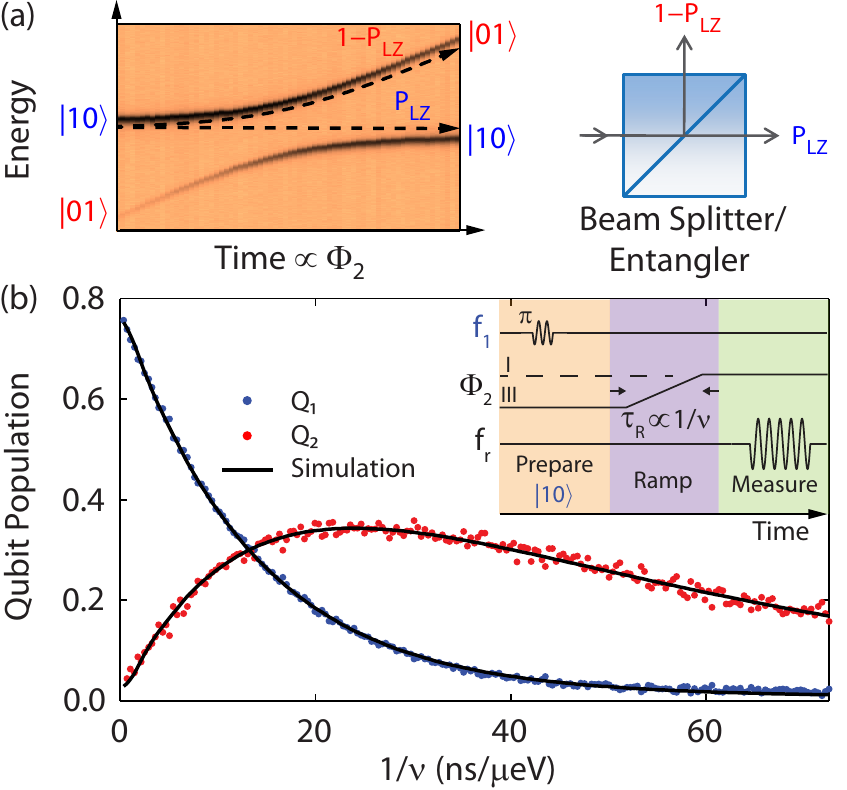}
\caption{\label{sc2} (color online) (a) Fast flux bias ramps are used to sweep the system through the cavity-mediated avoided crossing, with dynamics governed by the Landau-Zener transition formula. Passage through the avoided crossing is equivalent to a beam splitter. If the system is initialized in state $|10\rangle$ and then swept through the avoided crossing, the probability to remain in $|10\rangle$ is given by $P_{\rm LZ}$. (b) The cavity-mediated beam splitter is characterized by sweeping through the avoided crossing from point III to point I for a range of ramp times $\tau_{\text{R}}\propto 1/\nu$. $Q_1$ ($Q_2$) populations are plotted as a function of $1/\nu$ in blue (red).}
\end{center}
\vspace{-0.3cm}
\end{figure}

To probe the Landau-Zener transition dynamics associated with the avoided crossing, we perform a single-passage experiment in which the state $|10\rangle$ is initialized at point III and then ramped over a pulse risetime $\tau_{\text{R}}$ to point I, where the resulting qubit populations are measured. A schematic illustration of the process is shown in Fig.\ \ref{sc2}(a) and the specific pulse sequence is depicted in the inset of Fig.\ \ref{sc2}(b). The Landau-Zener formula predicts a non-adiabatic transition probability $P_\text{LZ} = e^{-\frac{2\pi\Delta^2}{\hbar\nu}}$, where $\nu$ is the ``level velocity" of the uncoupled energies, $\nu \equiv \left|d(E_1 - E_2)/dt\right|$ \cite{Shevchenko2010}. We plot the resulting populations as a function of $1/\nu \propto \tau_{\text{R}}$ in Fig.\ \ref{sc2}(b). For large level velocities, Q$_1$ remains in its initial state and Q$_2$ remains in the ground state. As $\nu$ decreases, $P_{\text{LZ}}$ decreases from 1, and population transfer occurs. We note that in the absence of relaxation, the population of $Q_2$ (red points) would continue to increase asymptotically with $\tau_{\text{R}}$ to a population of 1, corresponding to a completely adiabatic process. Selecting $\tau_{\text{R}} \approx 40$ ns ($1/\nu \approx 13$ ns/$\mu$eV) gives a 50-50 beam splitter. Deviations from $P_{\rm LZ}$ = 1/2 at this point are due to qubit relaxation during the experiment. Fitting an exponential decay to the $Q_1$ curve (blue points) according to the Landau-Zener formula (after subtracting out the independently measured $T_{1,1} = 120$ ns $Q_1$ relaxation rate) yields a best fit value $2g_{12}/2\pi = 34.0$ MHz, within 1 MHz of the value extracted from spectroscopy.
Our experimental results agree well with numerical simulations [see solid lines in Fig.\ \ref{sc2}(b)] of a master equation in the Lindblad form that accounts for relaxation and dephasing \cite{Lindblad1976,Bishop2010}
\begin{equation}
\dot{\rho}=\frac{1}{i\hbar}\left[H,\rho\right] +\sum_{k,Q_i}(L_{k,Q_i}\rho L_{k,Q_i}^\dagger - \frac{1}{2}\{\rho,L_{k,Q_i}^\dagger L_{k,Q_i}\}).\label{lindblad}
\end{equation}

\noindent The Lindblad operator $L_{1,Q_{1}} = \sqrt{\Gamma_{1,Q_{1}}}\sigma_-\otimes I$ describes relaxation of $Q_{1}$ to its ground state,
\begin{figure*}[t]
\begin{center}
\includegraphics[width=2\columnwidth]{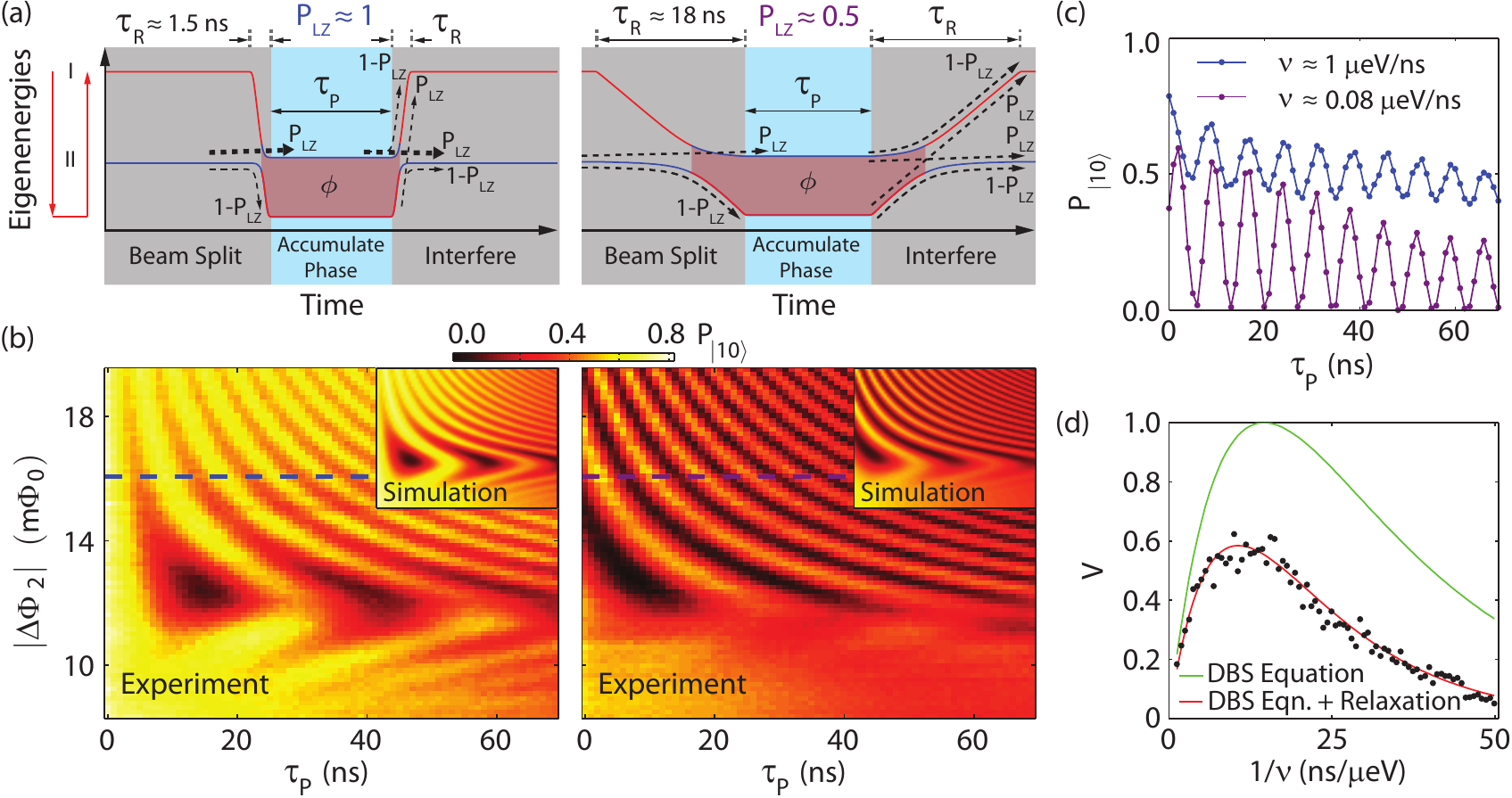}
\caption{\label{sc3} (color online) (a) A sweep back and forth across the avoided crossing is equivalent to a Mach-Zehnder interferometer, with level velocity-dependent transmission ratios that are controlled by the ramp time $\tau_{\text{R}}$ of the flux bias pulse. The red and blue curves depict how the eigenenergies evolve during the experiment. The labels I and II on the left refer to the labeled flux bias points in Fig.\ \ref{sc1}(b). (b) $P_{|10\rangle}$ plotted as a function of flux bias pulse width $\tau_{\text{P}}$ and flux bias pulse detuning $\Delta \Phi_2$ for ramp times $\tau_{\text{R}} \approx 1.5$ ns (left) and $\tau_{\text{R}} = 18$ ns (right). A pulse to the avoided crossing (point II) corresponds to $\left|\Delta \Phi_2\right|\approx 12$ m$\Phi_0$. Insets: Simulated Landau-Zener-St\"{u}ckelberg interference patterns. (c) $P_{|10\rangle}$ as a function of $\tau_{\text{P}}$ for $\tau_{\text{R}} \approx 1.5$ ns and $\tau_{\text{R}}$ = 18 ns, as extracted from the data sets in panel (b) (dashed lines).
(d) Oscillation visibility $V$ plotted as a function of $1/\nu$. The green solid line is the result of the double beam splitter (DBS) formula (\ref{DBSFormula}) after plugging in previously extracted parameters and neglecting relaxation. The solid red line is the result of multiplying the green curve by an exponential decay (with previously measured time constant $T_{1,1}\approx T_{1,2}$) with respect to $1/\nu \propto \tau_{\text{R}}$, with offset and scaling based on measured ramp and buffer times (see Supplemental Material), and agrees closely with a curve obtained using the numerical master equation simulation.}
\end{center}
\vspace{-0.4cm}
\end{figure*}
while $L_{2,Q_{1}} = \sqrt{\Gamma_{2,Q_{1}}}\sigma_z\otimes I$ describes pure dephasing of $Q_{1}$, with the analogous operators defined for $Q_2$. The weights of other potential Lindblad operators relevant to the dynamics were assumed to be small and were not included. The fitted rates $\Gamma$ are consistent with coherence times extracted from standard single-qubit Rabi and Ramsey experiments ($T_{1,1(2)} \approx 120(130)$ ns, $T_{2,1(2)}^* \approx 150(180)$ ns).

The Landau-Zener formula describes transition probabilities, but not the more fundamental transition \emph{amplitudes} that give rise to many interesting quantum interference phenomena. Phase is an important relation between two interacting states, in particular states that ``recombine" after a beam splitter event. If the avoided crossing is doubly traversed (across and back), the relative phase accumulated between the two consecutive crossings will lead to St\"{u}ckelburg oscillations \cite{Stuckelberg1932}. In conventional Landau-Zener problems such as atomic collisions, phase accumulation occurs so rapidly that interference is washed out by even a small amount of decoherence \cite{Stuckelberg1932}. However, in artificial atoms this phase can be observed and precisely tuned \cite{Oliver2005,Sillanpaa2006,Petta2010,Sun2010}.

In Fig.\ \ref{sc3} we observe St\"{u}ckelburg oscillations by performing a ``double beam splitter" experiment that is analogous to Mach-Zehnder interferometry. We are able to sensitively tune the visibility of the resulting St\"{u}ckelberg oscillations, in agreement with theoretical predictions. Schematics of a double passage experiment in the non-adiabatic ($P_\text{LZ} \approx 1$) and perfect beam splitter ($P_\text{LZ} \approx 0.5$) regimes are shown in Fig.\ \ref{sc3}(a). State $|10\rangle$ is prepared near point I and linear flux bias ramps are applied to sweep the system through the avoided crossing (point II) and back, with symmetric ramp time $\tau_{\text{R}}$ and pulse duration  $\tau_{\text{P}}$. The resulting experimental data are shown in Fig.\ \ref{sc3}(b), again with the left panel having $P_\text{LZ}\approx 1$ ($\tau_{\text{R}} \approx 1.5$ ns, leading to $\nu \approx 1$ $\mu$eV/ns and $P_\text{LZ}\approx 0.95$ at the dashed line) and right panel having $P_\text{LZ} \approx 0.5$ ($\tau_{\text{R}} = 18$ ns, leading to $\nu \approx 0.08$ $\mu$eV/ns and $P_\text{LZ}\approx 0.56$ at the dashed line). The resulting interference patterns agree well with the numerical master equation simulation (\ref{lindblad}), as shown in the insets of Fig.\ \ref{sc3}(b). Horizontal cuts through these data are shown in Fig.\ \ref{sc3}(c), indicating that the visibility of the oscillations is a sensitive function of the level velocity. Note that these oscillations occur in their entirety below $P_{|10\rangle}=1$ because of qubit relaxation during ramp and buffer times.

The level velocity-dependence of the oscillation visibility can be understood by examining the successive transition amplitudes during the double passage experiment. Neglecting relaxation and dephasing, each passage through the beam splitter performs an effective unitary operation on the incoming state according to the transfer matrix \cite{Shimshoni1991,Kayanuma1997,Sillanpaa2006}

\begin{equation}
 U =
 \begin{pmatrix}
  \sqrt{1-P_{\text{LZ}}}e^{i\tilde{\phi}_{\text{S}}} & i\sqrt{P_{\text{LZ}}}\\
  i\sqrt{P_{\text{LZ}}} & \sqrt{1-P_{\text{LZ}}}e^{-i\tilde{\phi}_{\text{S}}}
 \end{pmatrix},
\end{equation}
where $\tilde{\phi}_{\text{S}} = -\pi/2 + \phi_{\text{S}}$, $\phi_{\text{S}} = \pi/4 + \text{Arg}\left[\Gamma(1-i\delta)\right] + \delta(\ln \delta - 1)$ is the Stokes phase, and $\Gamma$ is the gamma function. The Stokes phase is a function of the adiabaticity parameter $\delta = \hbar g_{12}^2/\nu$ \cite{Sillanpaa2006}. Inserting a free evolution period between the two beam splitter operations results in a return probability
\begin{equation}
P_{|10\rangle} = 1 - 2 P_\text{LZ}(1-P_\text{LZ})\left[1+\cos(\phi - 2\tilde{\phi}_{\text{S}})\right],\label{DBSFormula}
\end{equation}
where $\phi$ is the dynamical phase accumulated on the far side of the avoided crossing. Therefore, in the absence of decoherence, the visibility of the Landau-Zener-St\"{u}ckelberg interference fringes is $V$ = $4P_\text{LZ}(1-P_\text{LZ})$. We explicitly test this dependence by measuring coherent oscillations as a function of $\tau_{\text{P}}$ at a fixed $\left|\Delta \Phi_2\right| \approx 16 $ m$\Phi_0$ for different beam splitter ramp times $\tau_{\text{R}}$. The resulting oscillation visibilities, defined as the initial amplitudes of fits to the decaying oscillations (see Supplemental Material), are plotted as a function of $1/\nu$ in Fig.\ \ref{sc3}(d). Our results agree well with theory when relaxation is taken into account and yield a maximum oscillation visibility when $1/\nu \approx 12$ ns/$\mu$eV, consistent with the single-pass beam splitter calibration in Fig.\ \ref{sc2}(b).

\begin{figure}[t]
\begin{center}
		\includegraphics[width=\columnwidth]{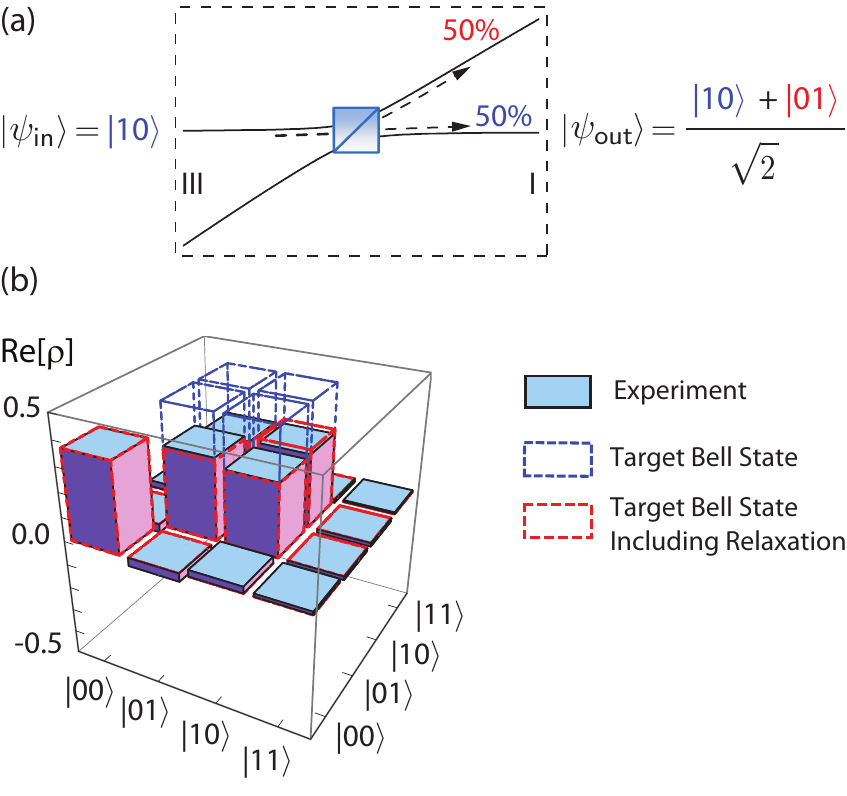}
\caption{\label{sc4} (color online) (a) In the 50-50 beam splitter limit, a single pass through the avoided crossing converts the input state $|\psi_{\text{in}}\rangle$ = $|10\rangle$ to the Bell state $|\psi_{\text{out}}\rangle$ = $1/{\sqrt{2}}\left(|10\rangle+|01\rangle\right) = |\psi_{\text{B}}\rangle$. (b) An entangled state is generated by preparing the system in state $|10\rangle$ and then sweeping through the avoided crossing with a ramp time corresponding to $1/\nu = 13$ ns/$\mu$eV and $P_{\text{LZ}} = 0.5$. We perform state tomography to extract the density matrix of the two-qubit system via maximum-likelihood estimation \cite{DiCarlo2009}, assuming a joint measurement operator of the form $M = \beta_{\text{II}} II + \beta_{\text{ZI}}ZI + \beta_{\text{IZ}}IZ + \beta_{\text{ZZ}}ZZ$ \cite{DiCarlo2010}. The density matrix shown is after an azimuthal $Z$ rotation on $Q_2$ chosen to maximize the off-diagonal terms in $\text{Re}\left[\rho\right]$. Imaginary parts (not shown) are all small compared to the real components (less than 0.04).  See Supplemental Material for further experimental details.}
\end{center}
\vspace{-0.4cm}
\end{figure}

Finally, we demonstrate entanglement generation through a single passage in the 50-50 beam splitter limit. After initializing the system in state $|10\rangle$ at point III, we sweep across the avoided crossing with a ramp rate corresponding to $P_\text{LZ} = 0.5$. This time, a set of tomography pre-rotations $U_{\rm k} \in SU(2)\otimes SU(2)$ are performed at point I directly before measurement. Figure \ref{sc4}(a) illustrates the ideal process in the absence of relaxation and dephasing. The initial state $|\psi_{\text{in}}\rangle$ = $|10\rangle$ is incident on the beam splitter, yielding an entangled state $|\psi_{\text{out}}\rangle$ = $1/{\sqrt{2}}$($|10\rangle$+$|01\rangle$). The results from state tomography are shown in Fig.\ \ref{sc4}(b) and agree well with the master equation simulation (\ref{lindblad}) that takes into account relaxation and dephasing (red dashed lines). We obtain a state fidelity with respect to the target Bell state of $F = \sqrt{\langle \psi_{\text{B}}|\rho|\psi_{\text{B}}\rangle} =$ 78\% and entanglement of formation $E_{\text{f}} = 63\%$ \cite{Wootters1998}, both limited by relaxation.

In conclusion, we have utilized cavity-mediated Landau-Zener physics to achieve coherent control of a two-qubit transmon system. We have verified the general Landau-Zener formula and the dependence of double-passage St\"{u}ckelberg oscillations on the level velocity. A simple entanglement generation protocol is demonstrated, where a single passage through the avoided crossing in the 50-50 beam splitter limit yields a Bell state with relaxation-limited $F$ = 78\%. The combination of single qubit rotations and Landau-Zener two-qubit interactions may be used to implement more complicated quantum algorithms in systems with longer coherence times.

\begin{acknowledgments}
Research was supported by the Sloan and Packard Foundations, DARPA QuEST grant HR0011-09-1-0007, and the NSF through the Princeton Center for Complex Materials (DMR-0819860) and Career program (DMR-0846341).
\end{acknowledgments}


\begin{thebibliography}{30}%
\makeatletter
\providecommand \@ifxundefined [1]{%
 \@ifx{#1\undefined}
}%
\providecommand \@ifnum [1]{%
 \ifnum #1\expandafter \@firstoftwo
 \else \expandafter \@secondoftwo
 \fi
}%
\providecommand \@ifx [1]{%
 \ifx #1\expandafter \@firstoftwo
 \else \expandafter \@secondoftwo
 \fi
}%
\providecommand \natexlab [1]{#1}%
\providecommand \enquote  [1]{``#1''}%
\providecommand \bibnamefont  [1]{#1}%
\providecommand \bibfnamefont [1]{#1}%
\providecommand \citenamefont [1]{#1}%
\providecommand \href@noop [0]{\@secondoftwo}%
\providecommand \href [0]{\begingroup \@sanitize@url \@href}%
\providecommand \@href[1]{\@@startlink{#1}\@@href}%
\providecommand \@@href[1]{\endgroup#1\@@endlink}%
\providecommand \@sanitize@url [0]{\catcode `\\12\catcode `\$12\catcode
  `\&12\catcode `\#12\catcode `\^12\catcode `\_12\catcode `\%12\relax}%
\providecommand \@@startlink[1]{}%
\providecommand \@@endlink[0]{}%
\providecommand \url  [0]{\begingroup\@sanitize@url \@url }%
\providecommand \@url [1]{\endgroup\@href {#1}{\urlprefix }}%
\providecommand \urlprefix  [0]{URL }%
\providecommand \Eprint [0]{\href }%
\providecommand \doibase [0]{http://dx.doi.org/}%
\providecommand \selectlanguage [0]{\@gobble}%
\providecommand \bibinfo  [0]{\@secondoftwo}%
\providecommand \bibfield  [0]{\@secondoftwo}%
\providecommand \translation [1]{[#1]}%
\providecommand \BibitemOpen [0]{}%
\providecommand \bibitemStop [0]{}%
\providecommand \bibitemNoStop [0]{.\EOS\space}%
\providecommand \EOS [0]{\spacefactor3000\relax}%
\providecommand \BibitemShut  [1]{\csname bibitem#1\endcsname}%
\let\auto@bib@innerbib\@empty
\bibitem [{\citenamefont {Cohen-Tannoudji}\ \emph {et~al.}(1977)\citenamefont
  {Cohen-Tannoudji}, \citenamefont {Diu},\ and\ \citenamefont
  {Lalo\"{e}}}]{Cohen-Tannoudji1977}%
  \BibitemOpen
  \bibfield  {author} {\bibinfo {author} {\bibfnamefont {C.}~\bibnamefont
  {Cohen-Tannoudji}}, \bibinfo {author} {\bibfnamefont {B.}~\bibnamefont
  {Diu}}, \ and\ \bibinfo {author} {\bibfnamefont {F.}~\bibnamefont
  {Lalo\"{e}}},\ }\href@noop {} {\emph {\bibinfo {title} {Quantum
  Mechanics}}},\ Vol.~\bibinfo {volume} {1}\ (\bibinfo  {publisher} {Wiley, New
  York},\ \bibinfo {year} {1977})\ Chap.\ \bibinfo {chapter} {4C.}\BibitemShut
  {Stop}%
\bibitem [{\citenamefont {Landau}(1932)}]{Landau1932}%
  \BibitemOpen
  \bibfield  {author} {\bibinfo {author} {\bibfnamefont {L.}~\bibnamefont
  {Landau}},\ }\href@noop {} {\bibfield  {journal} {\bibinfo  {journal} {Phys.
  Z. Sowjetunion}\ }\textbf {\bibinfo {volume} {2}},\ \bibinfo {pages} {46}
  (\bibinfo {year} {1932})}\BibitemShut {NoStop}%
\bibitem [{\citenamefont {Zener}(1932)}]{Zener1932}%
  \BibitemOpen
  \bibfield  {author} {\bibinfo {author} {\bibfnamefont {C.}~\bibnamefont
  {Zener}},\ }\href {\doibase 10.1098/rspa.1932.0165} {\bibfield  {journal}
  {\bibinfo  {journal} {Proc. R. Soc. A}\ }\textbf {\bibinfo {volume} {137}},\
  \bibinfo {pages} {696} (\bibinfo {year} {1932})}\BibitemShut {NoStop}%
\bibitem [{\citenamefont {St\"{u}ckelberg}(1932)}]{Stuckelberg1932}%
  \BibitemOpen
  \bibfield  {author} {\bibinfo {author} {\bibfnamefont {E.~C.~G.}\
  \bibnamefont {St\"{u}ckelberg}},\ }\href {\doibase 10.5169/seals-110177}
  {\bibfield  {journal} {\bibinfo  {journal} {Helv. Phys. Acta}\ }\textbf
  {\bibinfo {volume} {5}},\ \bibinfo {pages} {369} (\bibinfo {year}
  {1932})}\BibitemShut {NoStop}%
\bibitem [{\citenamefont {Majorana}(1932)}]{Majorana1932}%
  \BibitemOpen
  \bibfield  {author} {\bibinfo {author} {\bibfnamefont {E.}~\bibnamefont
  {Majorana}},\ }\href {\doibase 10.1007/BF02960953} {\bibfield  {journal}
  {\bibinfo  {journal} {Nuovo Cimento}\ }\textbf {\bibinfo {volume} {9}},\
  \bibinfo {pages} {43} (\bibinfo {year} {1932})}\BibitemShut {NoStop}%
\bibitem [{\citenamefont {Nakamura}(2002)}]{Nakamura2002}%
  \BibitemOpen
  \bibfield  {author} {\bibinfo {author} {\bibfnamefont {H.}~\bibnamefont
  {Nakamura}},\ }\href@noop {} {\emph {\bibinfo {title} {Nonadiabatic
  Transition}}}\ (\bibinfo  {publisher} {World Scientific},\ \bibinfo {year}
  {2002})\BibitemShut {NoStop}%
\bibitem [{\citenamefont {Shevchenko}\ \emph {et~al.}(2010)\citenamefont
  {Shevchenko}, \citenamefont {Ashhab},\ and\ \citenamefont
  {Nori}}]{Shevchenko2010}%
  \BibitemOpen
  \bibfield  {author} {\bibinfo {author} {\bibfnamefont {S.}~\bibnamefont
  {Shevchenko}}, \bibinfo {author} {\bibfnamefont {S.}~\bibnamefont {Ashhab}},
  \ and\ \bibinfo {author} {\bibfnamefont {F.}~\bibnamefont {Nori}},\ }\href
  {\doibase 10.1016/j.physrep.2010.03.002} {\bibfield  {journal} {\bibinfo
  {journal} {Phys. Rep.}\ }\textbf {\bibinfo {volume} {492}},\ \bibinfo {pages}
  {1} (\bibinfo {year} {2010})}\BibitemShut {NoStop}%
\bibitem [{\citenamefont {Izmalkov}\ \emph {et~al.}(2004)\citenamefont
  {Izmalkov}, \citenamefont {Grajcar}, \citenamefont {Il'ichev}, \citenamefont
  {Oukhanski}, \citenamefont {Wagner}, \citenamefont {Meyer}, \citenamefont
  {Krech}, \citenamefont {Amin}, \citenamefont {van~den Brink},\ and\
  \citenamefont {Zagoskin}}]{Izmalkov2004}%
  \BibitemOpen
  \bibfield  {author} {\bibinfo {author} {\bibfnamefont {A.}~\bibnamefont
  {Izmalkov}}, \bibinfo {author} {\bibfnamefont {M.}~\bibnamefont {Grajcar}},
  \bibinfo {author} {\bibfnamefont {E.}~\bibnamefont {Il'ichev}}, \bibinfo
  {author} {\bibfnamefont {N.}~\bibnamefont {Oukhanski}}, \bibinfo {author}
  {\bibfnamefont {T.}~\bibnamefont {Wagner}}, \bibinfo {author} {\bibfnamefont
  {H.~G.}\ \bibnamefont {Meyer}}, \bibinfo {author} {\bibfnamefont
  {W.}~\bibnamefont {Krech}}, \bibinfo {author} {\bibfnamefont {M.~H.~S.}\
  \bibnamefont {Amin}}, \bibinfo {author} {\bibfnamefont {A.~M.}\ \bibnamefont
  {van~den Brink}}, \ and\ \bibinfo {author} {\bibfnamefont {A.~M.}\
  \bibnamefont {Zagoskin}},\ }\href {\doibase doi:10.1209/epl/i2003-10200-6}
  {\bibfield  {journal} {\bibinfo  {journal} {Europhys. Lett.}\ }\textbf
  {\bibinfo {volume} {65}},\ \bibinfo {pages} {844} (\bibinfo {year}
  {2004})}\BibitemShut {NoStop}%
\bibitem [{\citenamefont {Oliver}\ \emph {et~al.}(2005)\citenamefont {Oliver},
  \citenamefont {Yu}, \citenamefont {Lee}, \citenamefont {Berggren},
  \citenamefont {Levitov},\ and\ \citenamefont {Orlando}}]{Oliver2005}%
  \BibitemOpen
  \bibfield  {author} {\bibinfo {author} {\bibfnamefont {W.~D.}\ \bibnamefont
  {Oliver}}, \bibinfo {author} {\bibfnamefont {Y.}~\bibnamefont {Yu}}, \bibinfo
  {author} {\bibfnamefont {J.~C.}\ \bibnamefont {Lee}}, \bibinfo {author}
  {\bibfnamefont {K.~K.}\ \bibnamefont {Berggren}}, \bibinfo {author}
  {\bibfnamefont {L.~S.}\ \bibnamefont {Levitov}}, \ and\ \bibinfo {author}
  {\bibfnamefont {T.~P.}\ \bibnamefont {Orlando}},\ }\href {\doibase
  10.1126/science.1119678} {\bibfield  {journal} {\bibinfo  {journal}
  {Science}\ }\textbf {\bibinfo {volume} {310}},\ \bibinfo {pages} {1653}
  (\bibinfo {year} {2005})}\BibitemShut {NoStop}%
\bibitem [{\citenamefont {Sillanp\"a\"a}\ \emph {et~al.}(2006)\citenamefont
  {Sillanp\"a\"a}, \citenamefont {Lehtinen}, \citenamefont {Paila},
  \citenamefont {Makhlin},\ and\ \citenamefont {Hakonen}}]{Sillanpaa2006}%
  \BibitemOpen
  \bibfield  {author} {\bibinfo {author} {\bibfnamefont {M.}~\bibnamefont
  {Sillanp\"a\"a}}, \bibinfo {author} {\bibfnamefont {T.}~\bibnamefont
  {Lehtinen}}, \bibinfo {author} {\bibfnamefont {A.}~\bibnamefont {Paila}},
  \bibinfo {author} {\bibfnamefont {Y.}~\bibnamefont {Makhlin}}, \ and\
  \bibinfo {author} {\bibfnamefont {P.}~\bibnamefont {Hakonen}},\ }\href
  {\doibase 10.1103/PhysRevLett.96.187002} {\bibfield  {journal} {\bibinfo
  {journal} {Phys. Rev. Lett.}\ }\textbf {\bibinfo {volume} {96}},\ \bibinfo
  {pages} {187002} (\bibinfo {year} {2006})}\BibitemShut {NoStop}%
\bibitem [{\citenamefont {Berns}\ \emph {et~al.}(2008)\citenamefont {Berns},
  \citenamefont {Rudner}, \citenamefont {Valenzuela}, \citenamefont {Berggren},
  \citenamefont {Oliver}, \citenamefont {Levitov},\ and\ \citenamefont
  {Orlando}}]{Berns2008}%
  \BibitemOpen
  \bibfield  {author} {\bibinfo {author} {\bibfnamefont {D.~M.}\ \bibnamefont
  {Berns}}, \bibinfo {author} {\bibfnamefont {M.~S.}\ \bibnamefont {Rudner}},
  \bibinfo {author} {\bibfnamefont {S.~O.}\ \bibnamefont {Valenzuela}},
  \bibinfo {author} {\bibfnamefont {K.~K.}\ \bibnamefont {Berggren}}, \bibinfo
  {author} {\bibfnamefont {W.~D.}\ \bibnamefont {Oliver}}, \bibinfo {author}
  {\bibfnamefont {L.~S.}\ \bibnamefont {Levitov}}, \ and\ \bibinfo {author}
  {\bibfnamefont {T.~P.}\ \bibnamefont {Orlando}},\ }\href {\doibase
  10.1038/nature07262} {\bibfield  {journal} {\bibinfo  {journal} {Nature}\
  }\textbf {\bibinfo {volume} {455}},\ \bibinfo {pages} {51} (\bibinfo {year}
  {2008})}\BibitemShut {NoStop}%
\bibitem [{\citenamefont {Sun}\ \emph {et~al.}(2010)\citenamefont {Sun},
  \citenamefont {Wen}, \citenamefont {Mao}, \citenamefont {Chen}, \citenamefont
  {Yu}, \citenamefont {Wu},\ and\ \citenamefont {Han}}]{Sun2010}%
  \BibitemOpen
  \bibfield  {author} {\bibinfo {author} {\bibfnamefont {G.}~\bibnamefont
  {Sun}}, \bibinfo {author} {\bibfnamefont {X.}~\bibnamefont {Wen}}, \bibinfo
  {author} {\bibfnamefont {B.}~\bibnamefont {Mao}}, \bibinfo {author}
  {\bibfnamefont {J.}~\bibnamefont {Chen}}, \bibinfo {author} {\bibfnamefont
  {Y.}~\bibnamefont {Yu}}, \bibinfo {author} {\bibfnamefont {P.}~\bibnamefont
  {Wu}}, \ and\ \bibinfo {author} {\bibfnamefont {S.}~\bibnamefont {Han}},\
  }\href {\doibase 10.1038/ncomms1050} {\bibfield  {journal} {\bibinfo
  {journal} {Nat. Commun.}\ }\textbf {\bibinfo {volume} {1}},\ \bibinfo {pages}
  {51} (\bibinfo {year} {2010})}\BibitemShut {NoStop}%
\bibitem [{\citenamefont {Petta}\ \emph {et~al.}(2010)\citenamefont {Petta},
  \citenamefont {Lu},\ and\ \citenamefont {Gossard}}]{Petta2010}%
  \BibitemOpen
  \bibfield  {author} {\bibinfo {author} {\bibfnamefont {J.~R.}\ \bibnamefont
  {Petta}}, \bibinfo {author} {\bibfnamefont {H.}~\bibnamefont {Lu}}, \ and\
  \bibinfo {author} {\bibfnamefont {A.~C.}\ \bibnamefont {Gossard}},\ }\href
  {\doibase 10.1126/science.1183628} {\bibfield  {journal} {\bibinfo  {journal}
  {Science}\ }\textbf {\bibinfo {volume} {327}},\ \bibinfo {pages} {669}
  (\bibinfo {year} {2010})}\BibitemShut {NoStop}%
\bibitem [{\citenamefont {Gaudreau}\ \emph {et~al.}(2011)\citenamefont
  {Gaudreau}, \citenamefont {Granger}, \citenamefont {Kam}, \citenamefont
  {Aers}, \citenamefont {Studenikin}, \citenamefont {Zawadzki}, \citenamefont
  {Pioro-Ladri{\`e}re}, \citenamefont {Wasilewski},\ and\ \citenamefont
  {Sachrajda}}]{Gaudreau2011}%
  \BibitemOpen
  \bibfield  {author} {\bibinfo {author} {\bibfnamefont {L.}~\bibnamefont
  {Gaudreau}}, \bibinfo {author} {\bibfnamefont {G.}~\bibnamefont {Granger}},
  \bibinfo {author} {\bibfnamefont {A.}~\bibnamefont {Kam}}, \bibinfo {author}
  {\bibfnamefont {G.~C.}\ \bibnamefont {Aers}}, \bibinfo {author}
  {\bibfnamefont {S.~A.}\ \bibnamefont {Studenikin}}, \bibinfo {author}
  {\bibfnamefont {P.}~\bibnamefont {Zawadzki}}, \bibinfo {author}
  {\bibfnamefont {M.}~\bibnamefont {Pioro-Ladri{\`e}re}}, \bibinfo {author}
  {\bibfnamefont {Z.~R.}\ \bibnamefont {Wasilewski}}, \ and\ \bibinfo {author}
  {\bibfnamefont {A.~S.}\ \bibnamefont {Sachrajda}},\ }\href {\doibase
  10.1038/NPHYS2149} {\bibfield  {journal} {\bibinfo  {journal} {Nature Phys.}\
  }\textbf {\bibinfo {volume} {8}},\ \bibinfo {pages} {54} (\bibinfo {year}
  {2011})}\BibitemShut {NoStop}%
\bibitem [{\citenamefont {Blais}\ \emph {et~al.}(2004)\citenamefont {Blais},
  \citenamefont {Huang}, \citenamefont {Wallraff}, \citenamefont {Girvin},\
  and\ \citenamefont {Schoelkopf}}]{Blais2004}%
  \BibitemOpen
  \bibfield  {author} {\bibinfo {author} {\bibfnamefont {A.}~\bibnamefont
  {Blais}}, \bibinfo {author} {\bibfnamefont {R.-S.}\ \bibnamefont {Huang}},
  \bibinfo {author} {\bibfnamefont {A.}~\bibnamefont {Wallraff}}, \bibinfo
  {author} {\bibfnamefont {S.~M.}\ \bibnamefont {Girvin}}, \ and\ \bibinfo
  {author} {\bibfnamefont {R.~J.}\ \bibnamefont {Schoelkopf}},\ }\href
  {\doibase 10.1103/PhysRevA.69.062320} {\bibfield  {journal} {\bibinfo
  {journal} {Phys. Rev. A}\ }\textbf {\bibinfo {volume} {69}},\ \bibinfo
  {pages} {062320} (\bibinfo {year} {2004})}\BibitemShut {NoStop}%
\bibitem [{\citenamefont {Wallraff}\ \emph {et~al.}(2004)\citenamefont
  {Wallraff}, \citenamefont {Schuster}, \citenamefont {Blais}, \citenamefont
  {Frunzio}, \citenamefont {Huang}, \citenamefont {Majer}, \citenamefont
  {Kumar}, \citenamefont {Girvin},\ and\ \citenamefont
  {Schoelkopf}}]{Wallraff2004}%
  \BibitemOpen
  \bibfield  {author} {\bibinfo {author} {\bibfnamefont {A.}~\bibnamefont
  {Wallraff}}, \bibinfo {author} {\bibfnamefont {D.~I.}\ \bibnamefont
  {Schuster}}, \bibinfo {author} {\bibfnamefont {A.}~\bibnamefont {Blais}},
  \bibinfo {author} {\bibfnamefont {L.}~\bibnamefont {Frunzio}}, \bibinfo
  {author} {\bibfnamefont {R.~S.}\ \bibnamefont {Huang}}, \bibinfo {author}
  {\bibfnamefont {J.}~\bibnamefont {Majer}}, \bibinfo {author} {\bibfnamefont
  {S.}~\bibnamefont {Kumar}}, \bibinfo {author} {\bibfnamefont {S.~M.}\
  \bibnamefont {Girvin}}, \ and\ \bibinfo {author} {\bibfnamefont {R.~J.}\
  \bibnamefont {Schoelkopf}},\ }\href {\doibase 10.1038/nature02851} {\bibfield
   {journal} {\bibinfo  {journal} {Nature}\ }\textbf {\bibinfo {volume}
  {431}},\ \bibinfo {pages} {162} (\bibinfo {year} {2004})}\BibitemShut
  {NoStop}%
\bibitem [{\citenamefont {Majer}\ \emph {et~al.}(2007)\citenamefont {Majer},
  \citenamefont {Chow}, \citenamefont {Gambetta}, \citenamefont {Koch},
  \citenamefont {Johnson}, \citenamefont {Schreier}, \citenamefont {Frunzio},
  \citenamefont {Schuster}, \citenamefont {Houck}, \citenamefont {Wallraff},
  \citenamefont {Blais}, \citenamefont {Devoret}, \citenamefont {Girvin},\ and\
  \citenamefont {Schoelkopf}}]{Majer2007}%
  \BibitemOpen
  \bibfield  {author} {\bibinfo {author} {\bibfnamefont {J.}~\bibnamefont
  {Majer}}, \bibinfo {author} {\bibfnamefont {J.~M.}\ \bibnamefont {Chow}},
  \bibinfo {author} {\bibfnamefont {J.~M.}\ \bibnamefont {Gambetta}}, \bibinfo
  {author} {\bibfnamefont {J.}~\bibnamefont {Koch}}, \bibinfo {author}
  {\bibfnamefont {B.~R.}\ \bibnamefont {Johnson}}, \bibinfo {author}
  {\bibfnamefont {J.~A.}\ \bibnamefont {Schreier}}, \bibinfo {author}
  {\bibfnamefont {L.}~\bibnamefont {Frunzio}}, \bibinfo {author} {\bibfnamefont
  {D.~I.}\ \bibnamefont {Schuster}}, \bibinfo {author} {\bibfnamefont {A.~A.}\
  \bibnamefont {Houck}}, \bibinfo {author} {\bibfnamefont {A.}~\bibnamefont
  {Wallraff}}, \bibinfo {author} {\bibfnamefont {A.}~\bibnamefont {Blais}},
  \bibinfo {author} {\bibfnamefont {M.~H.}\ \bibnamefont {Devoret}}, \bibinfo
  {author} {\bibfnamefont {S.~M.}\ \bibnamefont {Girvin}}, \ and\ \bibinfo
  {author} {\bibfnamefont {R.~J.}\ \bibnamefont {Schoelkopf}},\ }\href
  {\doibase 10.1038/nature06184} {\bibfield  {journal} {\bibinfo  {journal}
  {Nature}\ }\textbf {\bibinfo {volume} {449}},\ \bibinfo {pages} {443}
  (\bibinfo {year} {2007})}\BibitemShut {NoStop}%
\bibitem [{\citenamefont {Sillanp{\"a}{\"a}}\ \emph {et~al.}(2007)\citenamefont
  {Sillanp{\"a}{\"a}}, \citenamefont {Park},\ and\ \citenamefont
  {Simmonds}}]{Simmonds2007}%
  \BibitemOpen
  \bibfield  {author} {\bibinfo {author} {\bibfnamefont {M.~A.}\ \bibnamefont
  {Sillanp{\"a}{\"a}}}, \bibinfo {author} {\bibfnamefont {J.~I.}\ \bibnamefont
  {Park}}, \ and\ \bibinfo {author} {\bibfnamefont {R.~W.}\ \bibnamefont
  {Simmonds}},\ }\href {\doibase 10.1038/nature06124} {\bibfield  {journal}
  {\bibinfo  {journal} {Nature}\ }\textbf {\bibinfo {volume} {449}},\ \bibinfo
  {pages} {438} (\bibinfo {year} {2007})}\BibitemShut {NoStop}%
\bibitem [{\citenamefont {DiCarlo}\ \emph {et~al.}(2009)\citenamefont
  {DiCarlo}, \citenamefont {Chow}, \citenamefont {Gambetta}, \citenamefont
  {Bishop}, \citenamefont {Johnson}, \citenamefont {Schuster}, \citenamefont
  {Majer}, \citenamefont {Blais}, \citenamefont {Frunzio}, \citenamefont
  {Girvin},\ and\ \citenamefont {Schoelkopf}}]{DiCarlo2009}%
  \BibitemOpen
  \bibfield  {author} {\bibinfo {author} {\bibfnamefont {L.}~\bibnamefont
  {DiCarlo}}, \bibinfo {author} {\bibfnamefont {J.~M.}\ \bibnamefont {Chow}},
  \bibinfo {author} {\bibfnamefont {J.~M.}\ \bibnamefont {Gambetta}}, \bibinfo
  {author} {\bibfnamefont {L.~S.}\ \bibnamefont {Bishop}}, \bibinfo {author}
  {\bibfnamefont {B.~R.}\ \bibnamefont {Johnson}}, \bibinfo {author}
  {\bibfnamefont {D.~I.}\ \bibnamefont {Schuster}}, \bibinfo {author}
  {\bibfnamefont {J.}~\bibnamefont {Majer}}, \bibinfo {author} {\bibfnamefont
  {A.}~\bibnamefont {Blais}}, \bibinfo {author} {\bibfnamefont
  {L.}~\bibnamefont {Frunzio}}, \bibinfo {author} {\bibfnamefont
  {S.}~\bibnamefont {Girvin}}, \ and\ \bibinfo {author} {\bibfnamefont {R.~J.}\
  \bibnamefont {Schoelkopf}},\ }\href {\doibase 10.1038/nature08121} {\bibfield
   {journal} {\bibinfo  {journal} {Nature}\ }\textbf {\bibinfo {volume}
  {460}},\ \bibinfo {pages} {240} (\bibinfo {year} {2009})}\BibitemShut
  {NoStop}%
\bibitem [{\citenamefont {Reed}\ \emph {et~al.}(2012)\citenamefont {Reed},
  \citenamefont {DiCarlo}, \citenamefont {Nigg}, \citenamefont {Sun},
  \citenamefont {Frunzio}, \citenamefont {Girvin},\ and\ \citenamefont
  {Schoelkopf}}]{Reed2012}%
  \BibitemOpen
  \bibfield  {author} {\bibinfo {author} {\bibfnamefont {M.~D.}\ \bibnamefont
  {Reed}}, \bibinfo {author} {\bibfnamefont {L.}~\bibnamefont {DiCarlo}},
  \bibinfo {author} {\bibfnamefont {S.~E.}\ \bibnamefont {Nigg}}, \bibinfo
  {author} {\bibfnamefont {L.}~\bibnamefont {Sun}}, \bibinfo {author}
  {\bibfnamefont {L.}~\bibnamefont {Frunzio}}, \bibinfo {author} {\bibfnamefont
  {S.~M.}\ \bibnamefont {Girvin}}, \ and\ \bibinfo {author} {\bibfnamefont
  {R.~J.}\ \bibnamefont {Schoelkopf}},\ }\href {\doibase 10.1038/nature10786}
  {\bibfield  {journal} {\bibinfo  {journal} {Nature}\ }\textbf {\bibinfo
  {volume} {482}},\ \bibinfo {pages} {382} (\bibinfo {year}
  {2012})}\BibitemShut {NoStop}%
\bibitem [{\citenamefont {Koch}\ \emph {et~al.}(2007)\citenamefont {Koch},
  \citenamefont {Yu}, \citenamefont {Gambetta}, \citenamefont {Houck},
  \citenamefont {Schuster}, \citenamefont {Majer}, \citenamefont {Blais},
  \citenamefont {Devoret}, \citenamefont {Girvin},\ and\ \citenamefont
  {Schoelkopf}}]{Koch2007}%
  \BibitemOpen
  \bibfield  {author} {\bibinfo {author} {\bibfnamefont {J.}~\bibnamefont
  {Koch}}, \bibinfo {author} {\bibfnamefont {T.~M.}\ \bibnamefont {Yu}},
  \bibinfo {author} {\bibfnamefont {J.}~\bibnamefont {Gambetta}}, \bibinfo
  {author} {\bibfnamefont {A.~A.}\ \bibnamefont {Houck}}, \bibinfo {author}
  {\bibfnamefont {D.~I.}\ \bibnamefont {Schuster}}, \bibinfo {author}
  {\bibfnamefont {J.}~\bibnamefont {Majer}}, \bibinfo {author} {\bibfnamefont
  {A.}~\bibnamefont {Blais}}, \bibinfo {author} {\bibfnamefont {M.~H.}\
  \bibnamefont {Devoret}}, \bibinfo {author} {\bibfnamefont {S.~M.}\
  \bibnamefont {Girvin}}, \ and\ \bibinfo {author} {\bibfnamefont {R.~J.}\
  \bibnamefont {Schoelkopf}},\ }\href {\doibase 10.1103/PhysRevA.76.042319}
  {\bibfield  {journal} {\bibinfo  {journal} {Phys. Rev. A}\ }\textbf {\bibinfo
  {volume} {76}},\ \bibinfo {pages} {042319} (\bibinfo {year}
  {2007})}\BibitemShut {NoStop}%
\bibitem [{\citenamefont {Dolan}(1977)}]{Dolan1977}%
  \BibitemOpen
  \bibfield  {author} {\bibinfo {author} {\bibfnamefont {G.~J.}\ \bibnamefont
  {Dolan}},\ }\href {\doibase 10.1063/1.89690} {\bibfield  {journal} {\bibinfo
  {journal} {Appl. Phys. Lett.}\ }\textbf {\bibinfo {volume} {31}},\ \bibinfo
  {pages} {337} (\bibinfo {year} {1977})}\BibitemShut {NoStop}%
\bibitem [{\citenamefont {Schuster}\ \emph {et~al.}(2005)\citenamefont
  {Schuster}, \citenamefont {Wallraff}, \citenamefont {Blais}, \citenamefont
  {Frunzio}, \citenamefont {Huang}, \citenamefont {Majer}, \citenamefont
  {Girvin},\ and\ \citenamefont {Schoelkopf}}]{Schuster2005}%
  \BibitemOpen
  \bibfield  {author} {\bibinfo {author} {\bibfnamefont {D.~I.}\ \bibnamefont
  {Schuster}}, \bibinfo {author} {\bibfnamefont {A.}~\bibnamefont {Wallraff}},
  \bibinfo {author} {\bibfnamefont {A.}~\bibnamefont {Blais}}, \bibinfo
  {author} {\bibfnamefont {L.}~\bibnamefont {Frunzio}}, \bibinfo {author}
  {\bibfnamefont {R.-S.}\ \bibnamefont {Huang}}, \bibinfo {author}
  {\bibfnamefont {J.}~\bibnamefont {Majer}}, \bibinfo {author} {\bibfnamefont
  {S.~M.}\ \bibnamefont {Girvin}}, \ and\ \bibinfo {author} {\bibfnamefont
  {R.~J.}\ \bibnamefont {Schoelkopf}},\ }\href {\doibase
  10.1103/PhysRevLett.94.123602} {\bibfield  {journal} {\bibinfo  {journal}
  {Phys. Rev. Lett.}\ }\textbf {\bibinfo {volume} {94}},\ \bibinfo {pages}
  {123602} (\bibinfo {year} {2005})}\BibitemShut {NoStop}%
\bibitem [{\citenamefont {Reed}\ \emph {et~al.}(2010)\citenamefont {Reed},
  \citenamefont {DiCarlo}, \citenamefont {Johnson}, \citenamefont {Sun},
  \citenamefont {Schuster}, \citenamefont {Frunzio},\ and\ \citenamefont
  {Schoelkopf}}]{Reed2010}%
  \BibitemOpen
  \bibfield  {author} {\bibinfo {author} {\bibfnamefont {M.~D.}\ \bibnamefont
  {Reed}}, \bibinfo {author} {\bibfnamefont {L.}~\bibnamefont {DiCarlo}},
  \bibinfo {author} {\bibfnamefont {B.~R.}\ \bibnamefont {Johnson}}, \bibinfo
  {author} {\bibfnamefont {L.}~\bibnamefont {Sun}}, \bibinfo {author}
  {\bibfnamefont {D.~I.}\ \bibnamefont {Schuster}}, \bibinfo {author}
  {\bibfnamefont {L.}~\bibnamefont {Frunzio}}, \ and\ \bibinfo {author}
  {\bibfnamefont {R.~J.}\ \bibnamefont {Schoelkopf}},\ }\href {\doibase
  10.1103/PhysRevLett.105.173601} {\bibfield  {journal} {\bibinfo  {journal}
  {Phys. Rev. Lett.}\ }\textbf {\bibinfo {volume} {105}},\ \bibinfo {pages}
  {173601} (\bibinfo {year} {2010})}\BibitemShut {NoStop}%
\bibitem [{\citenamefont {Lindblad}(1976)}]{Lindblad1976}%
  \BibitemOpen
  \bibfield  {author} {\bibinfo {author} {\bibfnamefont {G.}~\bibnamefont
  {Lindblad}},\ }\href {\doibase 10.1007/BF01608499} {\bibfield  {journal}
  {\bibinfo  {journal} {Comm. Math. Phys.}\ }\textbf {\bibinfo {volume} {48}},\
  \bibinfo {pages} {119} (\bibinfo {year} {1976})}\BibitemShut {NoStop}%
\bibitem [{\citenamefont {Bishop}(2010)}]{Bishop2010}%
  \BibitemOpen
  \bibfield  {author} {\bibinfo {author} {\bibfnamefont {L.~S.}\ \bibnamefont
  {Bishop}},\ }\emph {\bibinfo {title} {Circuit Quantum Electrodynamics}},\
  \href {http://arxiv.org/abs/1007.3520} {Ph.D. thesis},\ \bibinfo  {school}
  {Yale University} (\bibinfo {year} {2010})\BibitemShut {NoStop}%
\bibitem [{\citenamefont {Shimshoni}\ and\ \citenamefont
  {Gefen}(1991)}]{Shimshoni1991}%
  \BibitemOpen
  \bibfield  {author} {\bibinfo {author} {\bibfnamefont {E.}~\bibnamefont
  {Shimshoni}}\ and\ \bibinfo {author} {\bibfnamefont {Y.}~\bibnamefont
  {Gefen}},\ }\href {\doibase 10.1016/0003-4916(91)90275-D} {\bibfield
  {journal} {\bibinfo  {journal} {Ann. Phys.}\ }\textbf {\bibinfo {volume}
  {210}},\ \bibinfo {pages} {16} (\bibinfo {year} {1991})}\BibitemShut
  {NoStop}%
\bibitem [{\citenamefont {Kayanuma}(1997)}]{Kayanuma1997}%
  \BibitemOpen
  \bibfield  {author} {\bibinfo {author} {\bibfnamefont {Y.}~\bibnamefont
  {Kayanuma}},\ }\href {\doibase 10.1103/PhysRevA.55.R2495} {\bibfield
  {journal} {\bibinfo  {journal} {Phys. Rev. A}\ }\textbf {\bibinfo {volume}
  {55}},\ \bibinfo {pages} {R2495} (\bibinfo {year} {1997})}\BibitemShut
  {NoStop}%
\bibitem [{\citenamefont {Wootters}(1998)}]{Wootters1998}%
  \BibitemOpen
  \bibfield  {author} {\bibinfo {author} {\bibfnamefont {W.~K.}\ \bibnamefont
  {Wootters}},\ }\href {\doibase 10.1103/PhysRevLett.80.2245} {\bibfield
  {journal} {\bibinfo  {journal} {Phys. Rev. Lett.}\ }\textbf {\bibinfo
  {volume} {80}},\ \bibinfo {pages} {2245} (\bibinfo {year}
  {1998})}\BibitemShut {NoStop}%
\bibitem [{\citenamefont {DiCarlo}\ \emph {et~al.}(2010)\citenamefont
  {DiCarlo}, \citenamefont {Reed}, \citenamefont {Sun}, \citenamefont
  {Johnson}, \citenamefont {Chow}, \citenamefont {Gambetta}, \citenamefont
  {Frunzio}, \citenamefont {Girvin}, \citenamefont {Devoret},\ and\
  \citenamefont {Schoelkopf}}]{DiCarlo2010}%
  \BibitemOpen
  \bibfield  {author} {\bibinfo {author} {\bibfnamefont {L.}~\bibnamefont
  {DiCarlo}}, \bibinfo {author} {\bibfnamefont {M.~D.}\ \bibnamefont {Reed}},
  \bibinfo {author} {\bibfnamefont {L.}~\bibnamefont {Sun}}, \bibinfo {author}
  {\bibfnamefont {B.~R.}\ \bibnamefont {Johnson}}, \bibinfo {author}
  {\bibfnamefont {J.~M.}\ \bibnamefont {Chow}}, \bibinfo {author}
  {\bibfnamefont {J.~M.}\ \bibnamefont {Gambetta}}, \bibinfo {author}
  {\bibfnamefont {L.}~\bibnamefont {Frunzio}}, \bibinfo {author} {\bibfnamefont
  {S.~M.}\ \bibnamefont {Girvin}}, \bibinfo {author} {\bibfnamefont {M.~H.}\
  \bibnamefont {Devoret}}, \ and\ \bibinfo {author} {\bibfnamefont {R.~J.}\
  \bibnamefont {Schoelkopf}},\ }\href {\doibase 10.1038/nature09416} {\bibfield
   {journal} {\bibinfo  {journal} {Nature}\ }\textbf {\bibinfo {volume}
  {467}},\ \bibinfo {pages} {574} (\bibinfo {year} {2010})}\BibitemShut
  {NoStop}%
\end{thebibliography}

\end{document}